\documentclass[]{aa}
\usepackage{graphicx}
\begin{document}
   \title{High resolution observations of SiO masers: comparing the spatial 
distribution at 43 and 86 GHz}

   \author{R. Soria-Ruiz\inst{1} \and J. Alcolea\inst{1} \and F. Colomer\inst{2} 
\and V. Bujarrabal\inst{2} \and J.-F. Desmurs\inst{1} \and K.B. Marvel\inst{3} \and \\P.J. Diamond\inst{4}
          }

   \offprints{{R.\,Soria-Ruiz}\\ \email{r.soria@oan.es}}

   \institute{Observatorio Astron\'omico Nacional, c/Alfonso XII 3, E-28014 Madrid, Spain \and
Observatorio Astron\'omico Nacional, Apartado 112, E-28803 Alcal\'a de Henares, Spain \and
American Astronomical Society, 2000 Florida Avenue NW Suite 400, Washington, DC 20009-1231, USA
 \and Jodrell Bank Observatory, University of Manchester, Macclesfield, Cheshire SK11 9DL, UK
    }

   \date{Received 22 April 2004 / Accepted 21 June 2004}

   \abstract{We present sub-milliarcsecond observations of SiO masers in the late-type 
stars IRC\,+10011 and $\chi$\,Cyg. We have used the NRAO Very Long Baseline Array (VLBA) 
to map the 43 GHz ($v$=1,\,2 $J$=1--0) and the 86 GHz ($v$=1,\,2 $J$=2--1) SiO masers. 
All the transitions have been imaged except the $v$=2 $J$=2--1 in IRC\,+10011. 
 We report the first VLBI map of the $v$=1 \mbox{$J$=2--1}
$^{28}$SiO maser in IRC\,+10011 as well as the first VLBA images of SiO masers in an 
S-type Mira variable, $\chi$\,Cyg. In this paper we have focused on the study of the relative spatial 
distribution of the different observed lines. We have found that in some cases 
the observational results are not reproduced by the current theoretical pumping models, either 
radiative or collisional. In particular, for IRC\,+10011, the $v$=1 $J$=1--0 and $J$=2--1 
$^{28}$SiO lines have different spatial distributions and 
emitting region sizes, the $J$=2--1 emission being located in an outer region of the envelope. 
For $\chi$\,Cyg, the distributions also differ, but the sizes of the masing regions are comparable.   
 We suggest that the line overlaps between ro-vibrational transitions of two abundant molecular
 species, H$_{2}$O and $^{28}$SiO, is a possible explanation for the discrepancies found 
between the observations and the theoretical predictions. 
We have introduced this overlapping process in the calculations of the 
excitation of the SiO molecule. We conclude that the line overlaps can strongly affect 
the excitation of SiO and may reproduce the unexpected observational results for the 
two sources studied.

   }

   \authorrunning{R. Soria-Ruiz et al.}
   \titlerunning{ VLBI multi-transition observations in circumstellar envelopes}

   \maketitle
%

\section{Introduction}

Circumstellar SiO maser emission is detected in Long Period Variables (AGB-stars): 
Mira-type, semi-regular and irregular variables, and OH/IR stars; of M and S spectral types. 
Due to the strong energy requirements for the pumping of these masers, they are produced in 
the innermost shells of the circumstellar envelopes around these stars, where the gas density
and stellar radiation field are important. On the other hand, this amplified emission is 
very bright and compact, therefore being an ideal target for Very Long Baseline 
Interferometric (VLBI) studies. In fact, VLBI observations are the only way to accurately map
 the structure of the SiO maser emitting regions, 
providing angular resolutions up to several 0.1 mas (Colomer et al. \cite{colomer}).

First estimations of the location of the SiO maser emission were derived from 
single-dish observations and from theoretical modelling (see e.g. Alcolea et al. 
\cite{alcolea0}; Bujarrabal et al. \cite{bujarrabal1}). 
These works already suggested that the SiO masers originate 
at distances of a few stellar radii from the central star, between the extended atmosphere 
and the dust-formation point. This has been confirmed in subsequent VLBI measurements 
(Danchi et al. \cite{danchi}; Greenhill et al. \cite{greenhill}; Wittkowski \& Boboltz 
\cite{wittkowski}) and in addition, that the distribution of the emission consists of 
ring-like structures composed of multiple maser spots with typical sizes of a few mas 
(Diamond et al. \cite{diamond1}; Greenhill et al. \cite{greenhill}; Desmurs et al. \cite{desmurs2}). 
VLBI monitoring of the SiO maser emission probes the kinematics of the gas in these
 inner shells, revealing inward/outward motions (Boboltz et al. \cite{boboltz1}; 
Diamond \& Kemball \cite{diamond2}) or showing indications of rotation of the maser 
shell in some peculiar objects (Boboltz \& Marvel \cite{boboltz2}; Hollis et al. 
\cite{hollis}; S\'anchez Contreras et al. \cite{sanchez}).


\begin{table*}[!ht]
\caption{Properties of the observed sources}
\begin{tabular}{ccccccccc}
\hline
\hline
&&&&&&&\\[-9pt]
source & R.A.& Dec.&variability&$\dot{M}$\,$^{\rm{a}}$&period\,$^{\rm{b}}$&phase\,$^{\rm{c}}$&dist.\,$^{\rm{d}}$&$V_{\rm{LSR}}$\,$^{\rm{e}}$\\
& (J2000)& (J2000)&type&($M_{\sun}$\,yr$^{-1}$)&(days)&&(kpc)&(km\,s$^{-1}$)\\[2pt]
 \hline
 \hline
&&&&&&&\\[-7pt]
IRC\,+10011 &01$^{\rm h}$06$^{\rm m}$25\fs99 & +12\degr35\arcmin53\farcs3&OH/IR&8.1\,$10^{-6}$&630&0.10&0.47&9.2\\[3pt]
$\chi$\,Cyg &19$^{\rm h}$50$^{\rm m}$33\fs92 & +32\degr54\arcmin50\farcs6&Mira &4.1\,$10^{-7}$&408&0.25&0.11&10.2\\[3pt]

\hline
\end{tabular}\\
[5pt]
$^{\rm{a}}$ Mass loss values derived following Loup et al. (\cite{loup}) but for the distances adopted here.\\
$^{\rm{b}}$ Periods form Pardo et al. (\cite{pardo}).\\
$^{\rm{c}}$ Stellar phases at the time of the observation: for IRC\,+10011, extrapolated value 
from Pardo et al. (\cite{pardo}) with respect to the NIR maximum (the typical phase lag between IR/SiO and 
optical maxima is $\sim$\,0.1 periods). For $\chi$\,Cyg, the optical phase derived from the 
light curve provided by the AAVSO (see http://www.aavso.org).\\
$^{\rm{d}}$ Distances from Herman et al. (\cite{herman1}) for IRC\,+10011 and from the HIPPARCOS catalogue 
for $\chi$\,Cyg.\\
$^{\rm{e}}$ Systemic velocity from high resolution CO observations by Cernicharo et al. (\cite{cernicharo}) 
and Bujarrabal et al. (\cite{bujarrabal3}) respectively.\\
\label{tab1}
\end{table*}


These observations also provide very valuable constraints for the theoretical models accounting 
for the origin and pumping of these masers, which are needed in order to understand the mechanism 
itself and to derive the physical conditions of the maser emitting regions from the observations. 
These theoretical models are divided in radiative and collisional, the 
main differences between both types being the main process that provides the energy to 
maintain the inversion of the population and therefore produce the maser transitions. 
The responsible mechanism in the case of the radiative models is the IR radiation from the 
central star (e.g.\ Bujarrabal \cite{bujarrabal4}, \cite{bujarrabal5}), whereas in the collisional 
models the energy is obtained from the kinetic energy of the gas (e.g.\ Lockett \& Elitzur 
\cite{lockett}). 

Most of the VLBI observational results on SiO masers have been obtained mainly in M-type Mira stars 
(simply because they typically present stronger maser emission than the other types of variables) and 
from observations of the 7\,mm rotational transitions of $^{28}$SiO: $v$=1 and 2 \mbox{$J$=1--0} lines.
 Successful VLBI observations of the 3\,mm $J$=2--1 rotational transition were made by 
Doeleman et al. (\cite{doeleman}) and Phillips \& Boboltz (\cite{phillips1}) towards the late type
 stars VX Sgr and $o$\,Cet, which helped in the study of the spatial structure of these 3\,mm SiO
 masers. Nowadays, using the NRAO\footnote{The National Radio Astronomy Observatory
 is a facility of the National Science Foundation operated under cooperative agreement
 by Associated Universities, Inc.} Very Long Baseline Array (VLBA), which was upgraded to 
operate at 3\,mm wavelength in 2001, it is possible to compare the observed results at 3\,mm 
with those at 7\,mm (Phillips et al. \cite{phillips2}).
 As we will see, the comparison of these quasi simultaneous observations with the theoretical
predictions for the $J$=2--1 lines will further constrain the models proposed to explain the 
physics involved in the SiO maser phenomenon.  

Making use of its new observational capabilities at 3\,mm, we have started 
multi-epoch/multi-transitional high resolution observations of SiO masers using the 
VLBA in a sample of long period variable stars. We have studied the 
SiO maser emission in the sources IRC\,+10011, $\chi$\,Cyg and TX\,Cam. In particular we have 
observed the $v$=1, $v$=2 $J$=1--0 and \mbox{$J$=2--1} transitions of $^{28}$SiO and the 
$v$=0 $J$=1--0 and $J$=2--1 lines of the $^{29}$SiO isotopic substitution (see Sect. 2). 

We present in this paper the results obtained in the first epoch for IRC\,+10011 and $\chi$\,Cyg 
(those for TX\,Cam will be presented in a separate paper). We report the first VLBI map of the 
$v$=1 \mbox{$J$=2--1} $^{28}$SiO maser in IRC\,+10011 as well as the first VLBA images of the 
$v$=1 and $v$=2 $J$=1--0 and $J$=2--1 transitions in an S-type Mira variable, $\chi$\,Cyg. 
We have detected all the observed lines except the $^{28}$SiO $v$=2 $J$=2--1 in IRC\,+10011 
and, unfortunately, the $^{29}$SiO lines in the two variables. 
A detailed description of the different maps obtained is given in Section 3.
 
In Section 4, we study and compare in detail the differences between the maser 
emissions at 7\,mm ($v$=1 and $v$=2 $J$=1--0) and between the 7 and 3\,mm ($v$=1 and 2, $J$=2--1 
and $J$=1--0). 
We focus on their relative spatial distribution, in order to compare the observational results with the 
predictions of the currently proposed pumping models for circumstellar SiO masers. We have found that 
our results for the 3 and 7\,mm comparison are not reproduced by any of these models, since all predict 
a similar emitting region at both wavelengths (Bujarrabal \cite{bujarrabal4}; Humphreys et al. 
\cite{humphreys2}) and we however obtain a totally different distribution in both IRC\,+10011 
and $\chi$\,Cyg. We propose the line overlaps between ro-vibrational transitions of two
abundant molecular species, H$_{2}$O and $^{28}$SiO, as a possible explanation of those unexpected 
results. For this reason, we have also introduced this overlapping process in the calculations and 
discuss the physical effects. Finally, the main conclusions and results derived from the high resolution 
observations and theoretical calculations performed are summarized in Section 5.


\begin{table*}
\caption{Observational setups}
\begin{center}
\begin{tabular}{cccccl}
&&&&&\\[-9pt]
\hline\hline
&&&&&\\[-9pt]
setup  & number & BW  & spectral  & spectral  & \qquad SiO transitions observed \\
       & of IFs       & (MHz)    & channels  & resolution (km/s)&(Left circular polarization only, except when noted)\\[2pt]
\hline\hline
&&&&&\\[-9pt]
7\,mm  &  4           & 8        & 256       & 0.22    &$^{28}$SiO $v$=1 $J$=1--0$^{\dag}$; $^{28}$SiO $v$=2 $J$=1--0; $^{29}$SiO $v$=0 $J$=1--0\\       
3\,mm-A&  2           & 16       & 512       & 0.11    &$^{28}$SiO $v$=2 $J$=2--1; $^{29}$SiO $v$=0 $J$=2--1\\  
3\,mm-B&  2           & 16       & 512       & 0.11    &$^{28}$SiO $v$=1 $J$=2--1$^{\dag}$\\ 
\hline \hline
\end{tabular}
\end{center}
$^{\dag}$ Left and right circular polarizations (LCP \& RCP) .\\
\label{tab2}
\end{table*}


\begin{table*}
\caption{Summary of the observational results}
\begin{center}
\begin{tabular}{|cccc|ccc|ccc|}
\cline{5-1}\cline{6-1}\cline{7-1}\cline{8-1}\cline{9-1}\cline{10-1}
\multicolumn{4}{c|}{}&&&&&&\\[-9pt]
\multicolumn{4}{c}{}&\multicolumn{3}{|c|}{IRC\,+10011}&\multicolumn{3}{c|}{$\chi$\,Cyg}\\
\hline\hline
&&&&&&&&&\\[-9pt]
  band&species &maser &rest frec. &restoring&\# of&masing region&restoring &\# of&masing region\\
& &transition &(MHz)& beam (mas)&spots&size (mas)&beam (mas)&spots&size (mas)\\[2pt]
\hline\hline
&&&&&&&&&\\[-9pt]
7\,mm & $^{28}$SiO& $v$=1 $J$=1--0 & 43122.080 & 1.1$\times$0.3  & 9  &19--24 & 0.7$\times$0.2 & 6 & 50--62 \\
 &        & $v$=2 $J$=1--0 & 42820.587 & 0.4$\times$0.2  & 11 &16--21 & 1.0$\times$0.6 & 1 & ---    \\
& $^{29}$SiO      & $v$=0 $J$=1--0 & 42879.916 & non-det.        & ---& ---   & non-det.       & --- &---  \\[2pt]
\hline
&&&&&&&&&\\[-9pt]
3\,mm &$^{28}$SiO & $v$=1 $J$=2--1 & 86243.442 &0.7$\times$0.3   & 8  &28--33 & 0.7$\times$0.5 & 6 & 50--58\\
 &        & $v$=2 $J$=2--1 & 85640.456 &non-det.         & ---&---    & 0.7$\times$0.3 & 4 & ---   \\
& $^{29}$SiO      & $v$=0 $J$=2--1 & 85759.132 &non-det.         & ---&---    & non-det.       &---& ---   \\[2pt]
 \hline
\end{tabular}
\end{center}
\label{tab3}
\end{table*}


\section{Observations and data analysis}

We have performed sub-milliarcsecond resolution observations of SiO masers transitions with the
 VLBA. We have observed the long period variable stars IRC\,+10011 on
 2002 January 5 and $\chi$\,Cyg on 2001 May 9 
(Table~\ref{tab1}). 
We have measured the $^{28}$SiO rotational lines $J$=1--0 and $J$=2--1 in the 
$v$=1 and $v$=2 vibrationally states and the $J$=1--0 and $J$=2--1 transitions of 
the $^{29}$SiO isotopic substitution in the ground vibrational state $v$=0.

All available antennas were used at the time of the observations, the entire array (10 antennas) 
for the 7\,mm (43 GHz) and 7 antennas for the 3\,mm (86 GHz) observations. 
The distances between the VLBA antennas range between 236\,km (LA-PT), and 8611\,km (SC-MK) for the 7\,mm
and 6156\,km (NL-MK) for the 3\,mm observations, thus achieving maximum angular resolutions of about 0.17 
and 0.12 mas respectively. 

The data were correlated at the VLBA correlator in Socorro (New Mexico, USA). The observations 
for each of the sources consisted of 3 different lines setups, the 7\,mm, the 3\,mm-A and 
3\,mm-B, which comprised a total observing time of about 8\,h per source. A summary of the observing and
correlator setups is presented in Table~\ref{tab2}. The $u$-$v$ coverages are shown in Fig.\,\ref{uv}.

The observations carried out are summarized in Table~\ref{tab3}. All the $^{28}$SiO transitions
 observed were detected except the $v$=2 $J$=2--1 maser in IRC\,+10011 and, unfortunately, none 
of the $^{29}$SiO lines because their flux was below the detection limit (they 
were not detected in the auto-correlations either). 

Using the mentioned setups, the four 7\,mm masers were observed simultaneously and the 7 and 3\,mm 
separated only by a few hours. This is important since one of the characteristics of SiO masers 
is their time variability. From monitoring programs of Mira and other types of regular 
variables, two types of variability have been reported: long-term variability on time scales 
of several months (see Alcolea et al. \cite{alcolea} and references therein), and changes 
on shorter time scales, 10-20 days (Pijpers et al. \cite{pijpers}). 
So far no significant variability has been observed on time scales 
of hours, and therefore our maps of the 7 and 3\,mm masers can be directly compared.

The data reduction was performed using the Astronomical Image Processing System (AIPS) package. 
 The calibrator used to calculate the bandpass and single-band delay corrections, 
which was later applied to the spectral line sources, was 3C\,454.3 for both sources.

The amplitude calibration was done by comparing a calibrated total power spectrum of the line source, 
used as a template, with the autocorrelations measured at each antenna, except in 
the $v$=1 $J$=\mbox{2--1} maser in IRC\,+10011 and in the $v$=2 $J$=2--1 transition in 
$\chi$\,Cyg. In these cases, we used the system temperatures and antenna gains tables provided 
by the NRAO within the data files.

The determination of the phase calibration was done following the standard method for spectral line 
VLBI data. 
Residual phase errors were removed later in a self-calibration process. 
 Finally, the integrated intensity maps were obtained independently for each of the SiO 
transitions detected, using the image deconvolution CLEAN algorithm.

Note that when the phase calibration is done by means of phase-referencing techniques, the absolute
 positions in the maps are lost and only relative positions to the reference 
channel are obtained. This adds a difficulty when comparing two maps since their relative 
position is unknown.

The results obtained are presented in Figs.\,2--7, where for each of the mapped SiO transitions,
 we show: \textbf{1)} The VLBA integrated intensity map (in Jy per clean beam). 
\textbf{2)} The individual spectra of all the maser features with a signal-to-noise ratio 
larger than 5 and appearing on at least three consecutive spectral channels. The spectrum 
for each spot was calculated transforming the map units (Jy/beam) to flux units (Janskys) 
for the selected maser region. \textbf{3)} The total power calibrated spectrum of the maser transition.
\textbf{4)} The total flux recovered in each frequency channel after the calibration
 and imaging. \textbf{5)} The ratio of the total power to the recovered flux.

   \begin{figure*}[!ht]
   \centering
\vspace{0.5cm} 
      \caption{$u$-$v$ coverage of the VLBA observations for IRC\,+10011 (upper panels) and $\chi$\,Cyg
 (lower panels) for the different frequency setups.}       
       \label{uv}
   \end{figure*} 

   \begin{figure*}[!ht]
   \centering
 \includegraphics[angle=-90, width=0.9\textwidth]{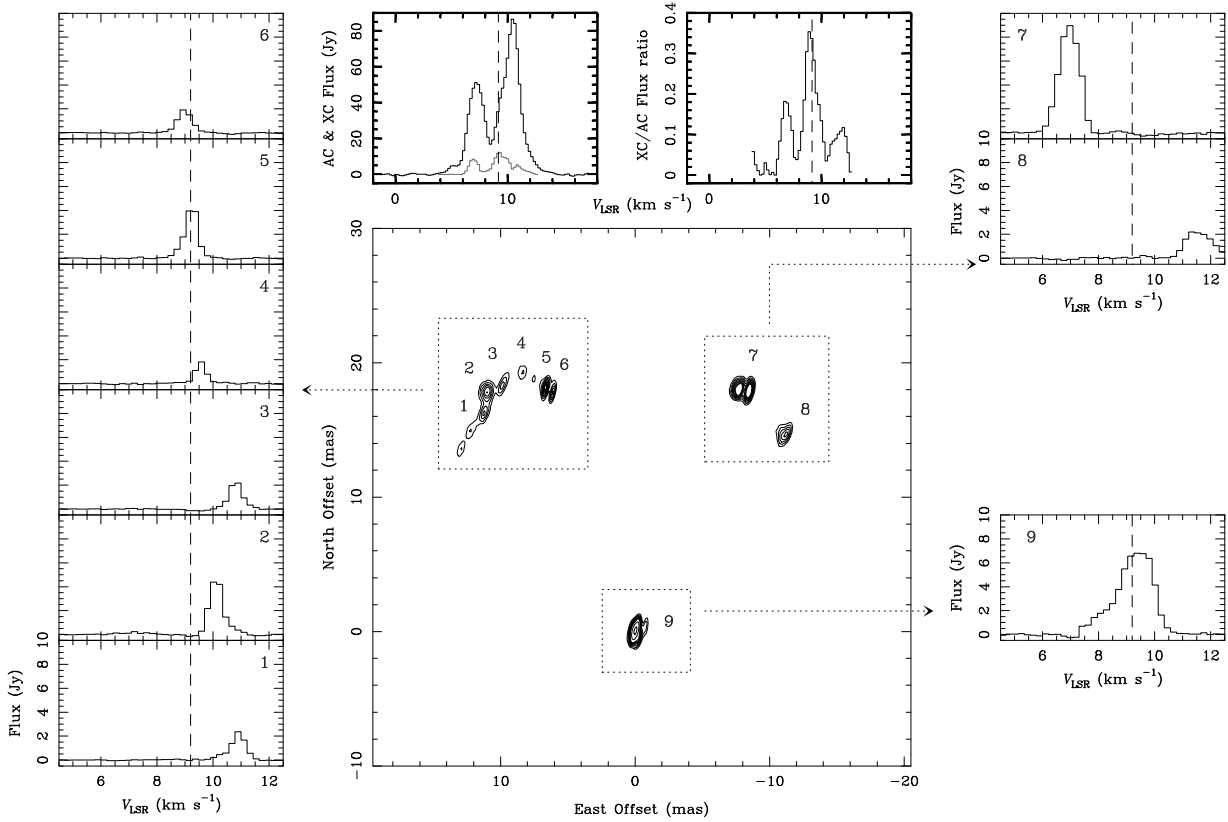}
      \caption{The $v$=1 $J$=1--0 $^{28}$SiO transition for IRC\,+10011. The figure 
consists of the integrated intensity map in Jy beam$^{-1}$$\cdot$\,km\,s$^{-1}$ units (center panel), 
the spectrum of the maser spots that appear on the map (numbered boxes), the total (AC)
 and recovered flux (XC) spectra (left panel above the map), and their ratio (right panel
 above the map). In all the spectra we indicate the systemic velocity with a dashed line. The peak
intensity of the map is 7.21 Jy beam$^{-1}$$\cdot$\,km\,s$^{-1}$ and the contour 
levels are 0.26, 0.44, 0.62, 0.80, 0.98, 1.16, 1.34, 1.52, 3.04 and 
6.08 Jy beam$^{-1}$$\cdot$\,km\,s$^{-1}$.}       
       \label{IRCv1}
   \end{figure*}

\section{Results}
In this section, we briefly describe the most relevant aspects of the observational results 
obtained for the $^{28}$SiO masers observed in the long-period variables IRC\,+10011 and $\chi$\,Cyg.

 We note that there is an additional difficulty when estimating the angular extents of the
 different masing regions since, in some cases, the distributions are not complete rings.
For this reason, we give an approximate range of the angular sizes derived from the 
brightest components that appear in the maps.

\subsection{IRC\,+10011}
IRC\,+10011 (WX Psc) is an OH/IR star of spectral type M9, characterized by a very thick 
circumstellar envelope and higher mass loss rate than Mira-type variable
 stars (see Table \ref{tab1}). The SiO emission of IRC\,+10011 varies in time with a period 
of about 630 days (Herman \& Habing \cite{herman2}; Pardo et al. \cite{pardo}); our observations
were performed close to a maximum.
All the $^{28}$SiO masers observed were detected except the $v$=2 $J$=2--1, probably due 
to the weak emission of this line in oxygen-rich variable stars (see Sect. 4.5).

\subsubsection{The $v$=1 $J$=1--0 transition}
The $v$=1 $J$=1--0 SiO results are displayed in Fig.\,\ref{IRCv1}. The intensity maps
 of the LCP and RCP observations were independently produced, although the same template
 autocorrelation spectra and phase reference channel (corresponding to a velocity 
of 9.16\,km\,s$^{-1}$) were used in both cases to calibrate the amplitude and the residual 
fringe-rates. No difference was found between the two circular polarizations and so the 
two maps were finally averaged. 

The interferometric VLBA map shows nine maser spots distributed in three different regions 
with velocities ranging from 6 to 12\,km\,s$^{-1}$ and peak intensities up to 9 Jy (spot 7). The
 angular diameter of the emitting region is $\sim$\,19--24 mas, derived from the brightest 
features. Almost 40\% of the flux has been recovered in the interferometric maps at 
velocities near the systemic, 9.2\,km\,s$^{-1}$.

   \begin{figure*}[!ht]
   \centering
   \includegraphics[angle=-90, width=0.9\textwidth]{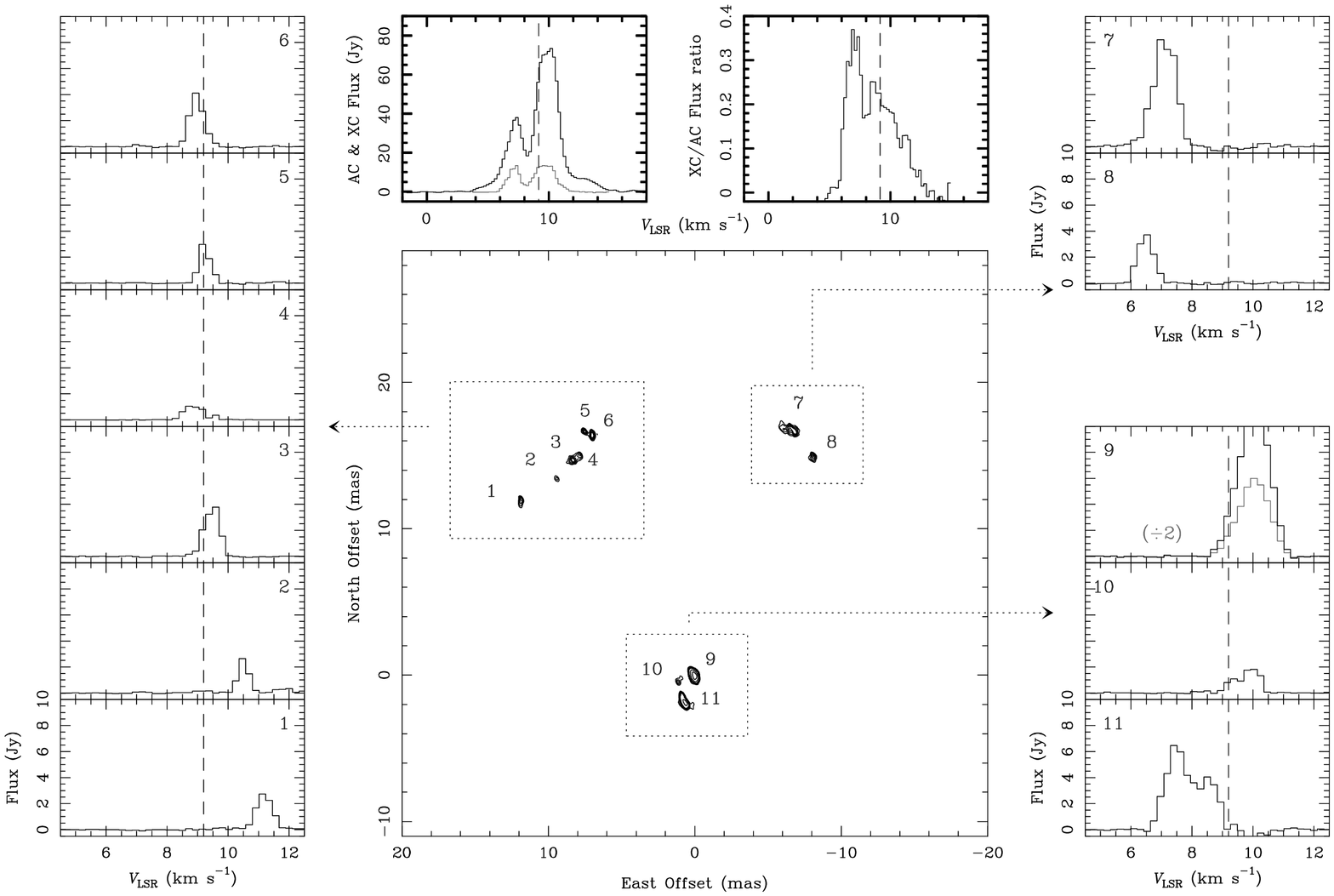}
   \caption{Same as Fig.\,\ref{IRCv1} for the $v$=2 $J$=1--0 SiO transition in IRC\,+10011. 
The peak intensity of the map is 5.74 Jy beam$^{-1}$$\cdot$\,km\,s$^{-1}$ and the 
contour levels are 0.35, 0.49, 0.63, 0.77, 0.91, 1.82 and 3.64 Jy beam$^{-1}$$\cdot$\,km\,s$^{-1}$.}
   \label{IRCv2}
   \end{figure*}

\subsubsection{The $v$=2 $J$=1--0 transition}
The $v$=2 $J$=1--0 results for IRC\,+10011 are presented in Fig.\,\ref{IRCv2}.
 The intensity map is composed of eleven maser clumps distributed into three different regions.
 The velocity range varies from 6 to 12\,km\,s$^{-1}$ and the intensities from $\sim$\,1 Jy 
(spot 4) to $\sim$\,12 Jy (spot 9, where the intensity has been divided by two in the plot 
to ease the comparison with the rest of the spectra). For this transition, the diameter of 
the emission is about \mbox{16--21 mas}. At velocities around 7\,km\,s$^{-1}$, about 40\% of the 
maser flux has been recovered.

\subsubsection{The $v$=1 $J$=2--1 transition}
\begin{figure*}[!ht]
   \centering
\includegraphics[angle=-90, width=0.9\textwidth]{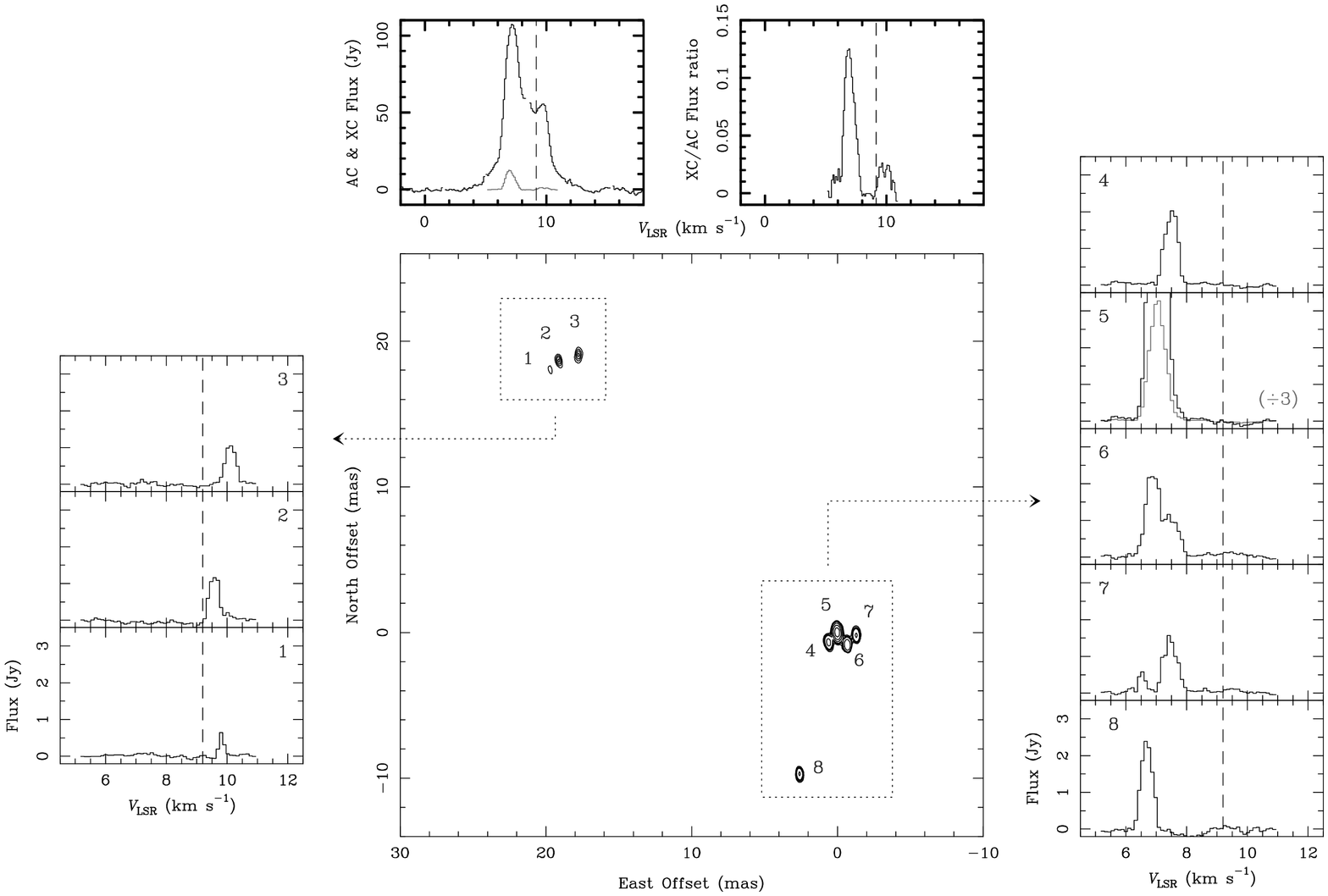}
   \caption{Same as Fig.\,\ref{IRCv1} for the $v$=1 $J$=2--1 SiO transition in IRC\,+10011. 
The peak intensity of the map is 5.71 Jy beam$^{-1}$$\cdot$\,km\,s$^{-1}$ and the contour 
levels are 0.12 (only in spots 1 to 3), 0.20, 0.28, 0.36, 0.44, 0.88, 1.76 and 3.52 Jy beam$^{-1}$$\cdot$\,km\,s$^{-1}$.}
   \label{IRC3mm}
   \end{figure*}

The $v$=1 $J$=2--1 SiO line is presented in Fig.\,\ref{IRC3mm}. We found eight
 different maser spots, with velocities from 6 to \mbox{10.5\,km\,s$^{-1}$} and intensities up to 
11 Jy (spot \mbox{number 9).} In this transition, about 13\% of the emission has been 
recovered at velocities $\sim$\,7\,km\,s$^{-1}$. The maser components are distributed in 
two regions with an angular separation of about 28--33 mas.  

\subsection{$\chi$\,Cyg} 

$\chi$\,Cyg is a Mira variable star with a mass loss rate similar to that typically found 
in other regular variables, and a period of about 408 days. This object was also observed 
near an SiO maximum \mbox{($\phi$\,=\,0.1--0.15, Alcolea et al. \cite{alcolea})};
 however, we note that in this case the variability of the SiO maser emission is not
as regular as in other Mira variables (Pardo et al. \cite{pardo}). $\chi$\,Cyg is an S7--9 spectral 
type star, thus characterized by a peculiar chemical composition. 
S-type variables have similar Carbon and Oxygen abundances, resulting in 
a lower production of \mbox{O-bearing} molecules other than CO. In particular, for $\chi$\,Cyg, 
the C/O ratio is found to be $\sim$\,0.95 (Keenan \& Boeshaar \cite{keenan}), a value relatively
low for an S-type star. 
Nevertheless, in this source the H$_{2}$O abundance might be significantly lower compared to
O-rich envelopes, since no H$_{2}$O maser emission has ever been detected.
 This aspect is relevant when explaining the pumping of SiO masers, since, as we will see in 
Section 4.5, the overlap between infrared lines of SiO and H$_{2}$O strongly affects the excitation 
of the former species.

\subsubsection{The $v$=1 $J$=1--0 transition}

\begin{figure*}[!ht]
   \centering
 \includegraphics[angle=-90, width=0.9\textwidth]{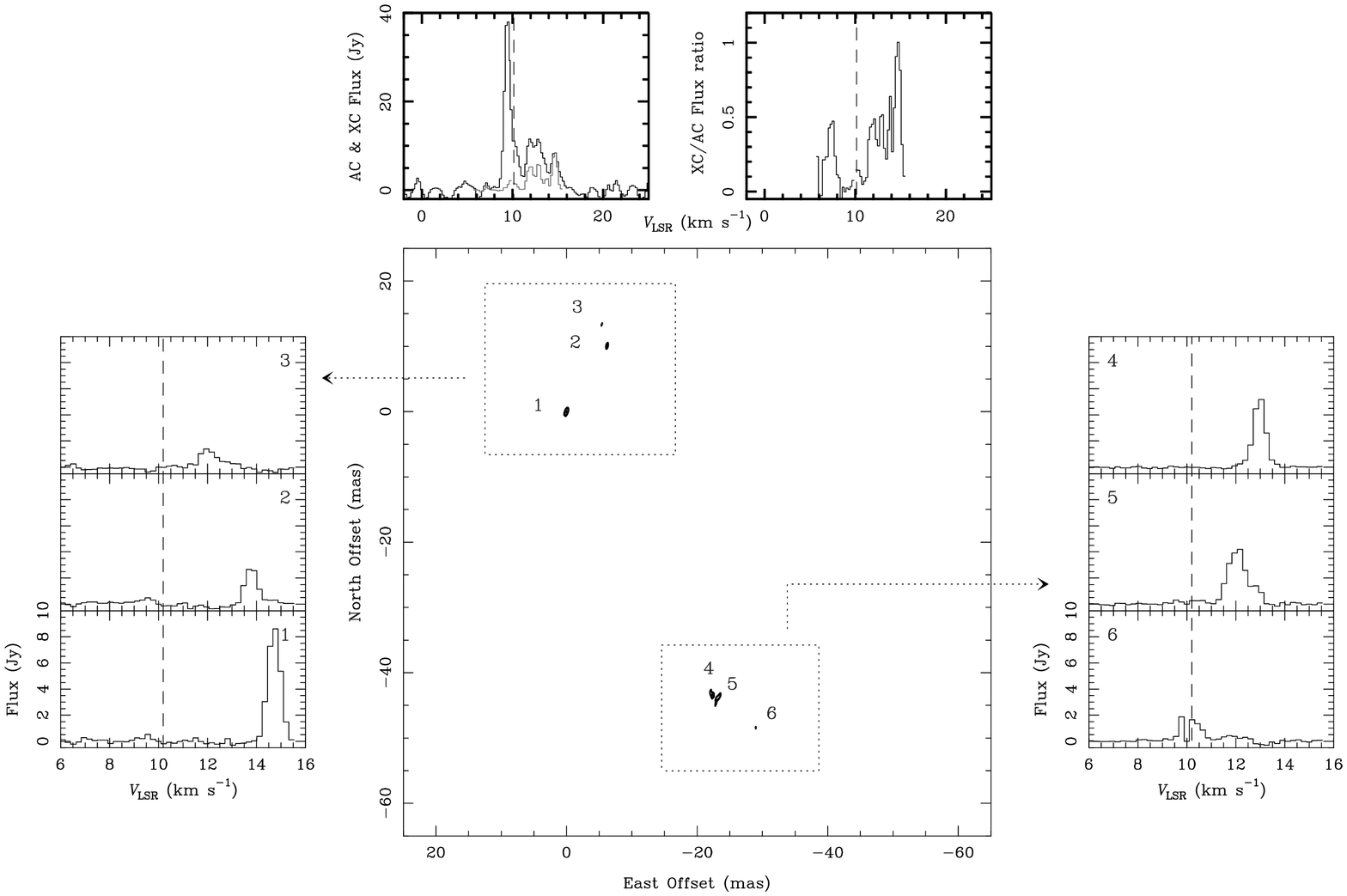}
      \caption{Same as Fig.\,\ref{IRCv1} for the $v$=1 $J$=1--0 SiO transition in $\chi$\,Cyg. 
The peak intensity of the map is 3.03 Jy beam$^{-1}$$\cdot$\,km\,s$^{-1}$ and the contour levels are 0.29,
0.39, 0.49, 0.68, 0.88, 1.08, 1.27 and 2.54 Jy beam$^{-1}$$\cdot$\,km\,s$^{-1}$.}
      \label{chiv1}
   \end{figure*}

The $v$=1 $J$=1--0 SiO maser for $\chi$\,Cyg is presented in Fig.\,\ref{chiv1}.
 Both, LCP and RCP observational data were averaged, after the calibration to produce
 the final map, since no significant differences were found between the two polarizations. 
Six maser spots appear in the map, with velocities from 9.8\,km\,s$^{-1}$ to 14\,km\,s$^{-1}$,
 distributed in two regions separated about 50--62 mas. The brightest feature 
\mbox{(spot 1)} has an intensity of about 8 Jy. All the flux has been recovered at velocities
 near 15\,km\,s$^{-1}$, and an average of 50\% for the other velocity channels.  

\subsubsection{The $v$=2 $J$=1--0 transition}

\begin{figure*}[!ht]
\hspace{0.7cm} 
 \includegraphics[angle=-90, width=0.67\textwidth]{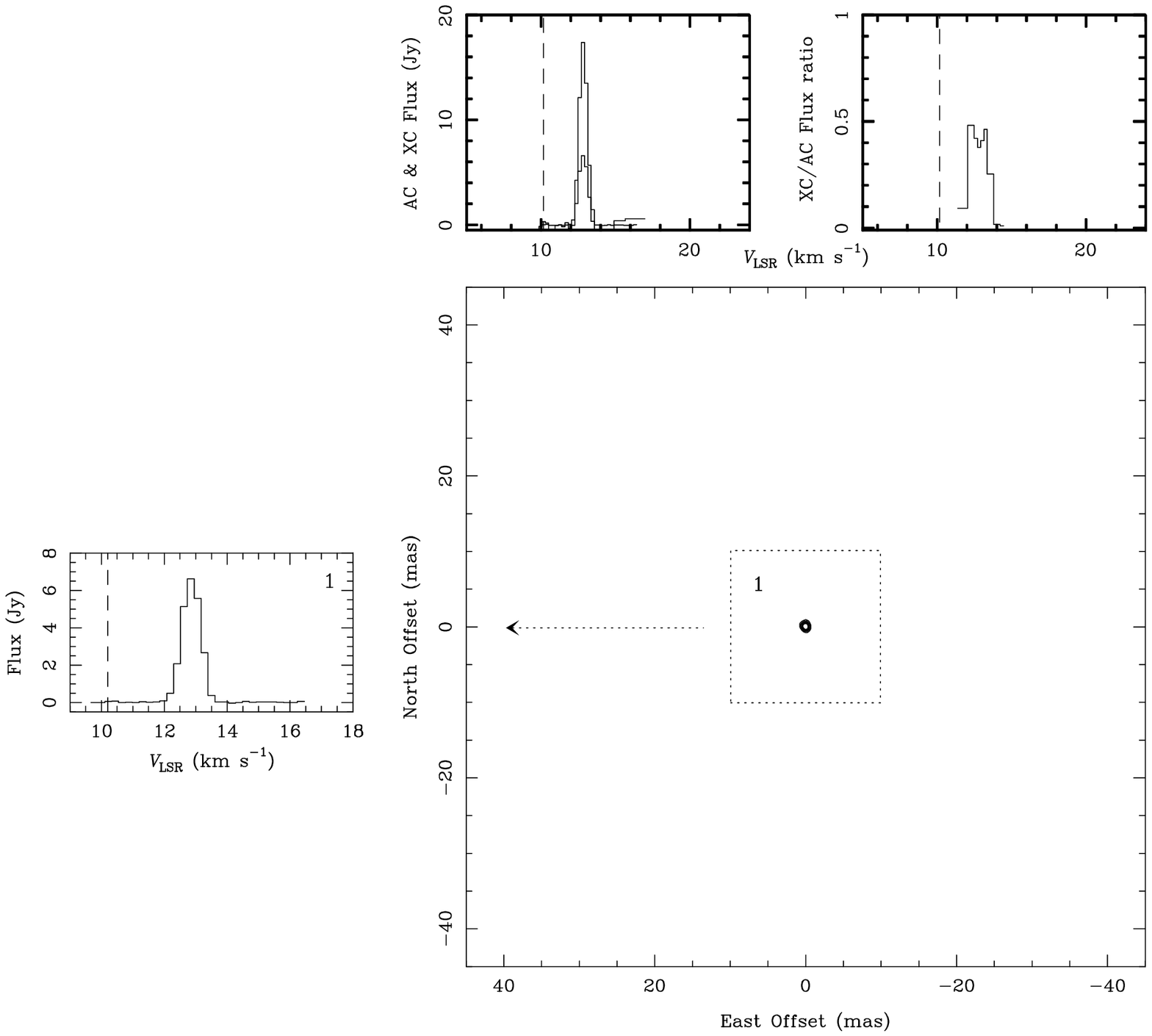}
      \caption{Same as Fig.\,\ref{IRCv1} for the $v$=2 $J$=1--0 SiO transition in $\chi$\,Cyg.
 The peak intensity of the map is 3.19 Jy beam$^{-1}$$\cdot$\,km\,s$^{-1}$ and the contour levels are 0.59,
0.98, 1.37, 1.76 and 2.15 Jy beam$^{-1}$$\cdot$\,km\,s$^{-1}$.}
      \label{chiv2}
   \end{figure*}

The $v$=2 $J$=1--0 transition is shown in Fig.\,\ref{chiv2}. 
Only one spot was found in the map, which is consistent with the simple line profile. Its velocity
 spread is \mbox{12--14\,km\,s$^{-1}$,} and its intensity is up to 8 Jy. 
Almost 50\% of the emission has been recovered.

\subsubsection{The $v$=1 $J$=2--1 transition}

\begin{figure*}[!ht]
 \centering 
 \includegraphics[angle=-90, width=0.9\textwidth]{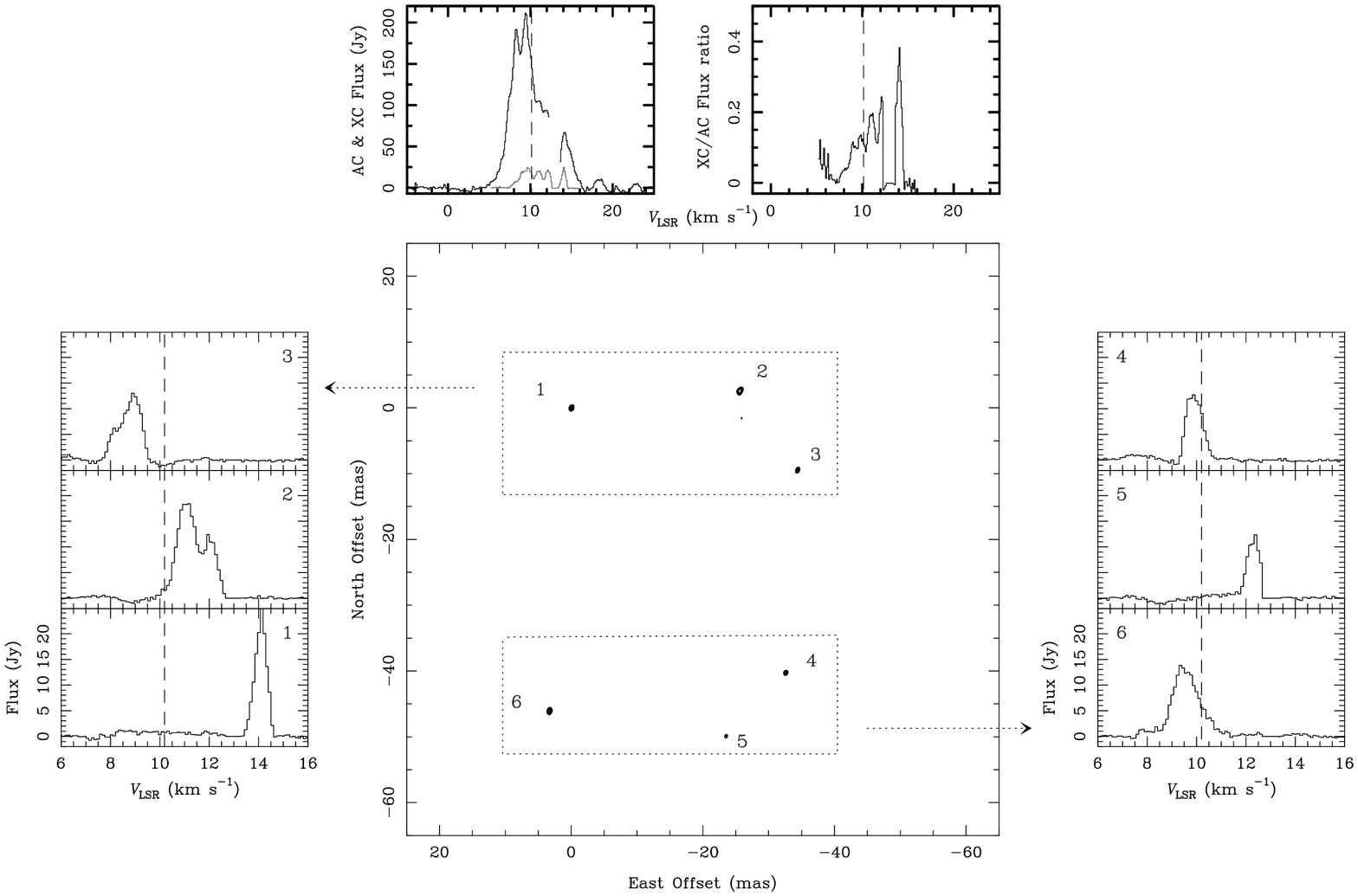}
      \caption{Same as Fig.\,\ref{IRCv1} for the $v$=1 $J$=2--1 SiO transition in $\chi$\,Cyg. 
The peak intensity of the map is 18.36 Jy beam$^{-1}$$\cdot$\,km\,s$^{-1}$ and the contour levels are 3.80,
 5.43, 7.06, 8.69, 11.95 and 13.58 Jy beam$^{-1}$$\cdot$\,km\,s$^{-1}$.}
         \label{chi3mm2}
   \end{figure*}

The $v$=1 $J$=2--1 SiO maser is presented in Fig.\,\ref{chi3mm2}. The VLBI map has six maser 
spots delineating a ring with a diameter of 50--58 mas, and with velocities from 10 to 
15\,km\,s$^{-1}$. The brightest spot has a flux density of about 25 Jy. Only 20\% of the emission was 
recovered in most velocity channels. 

  \begin{figure*}[!ht]
   \centering
 \includegraphics[angle=-90, width=0.85\textwidth]{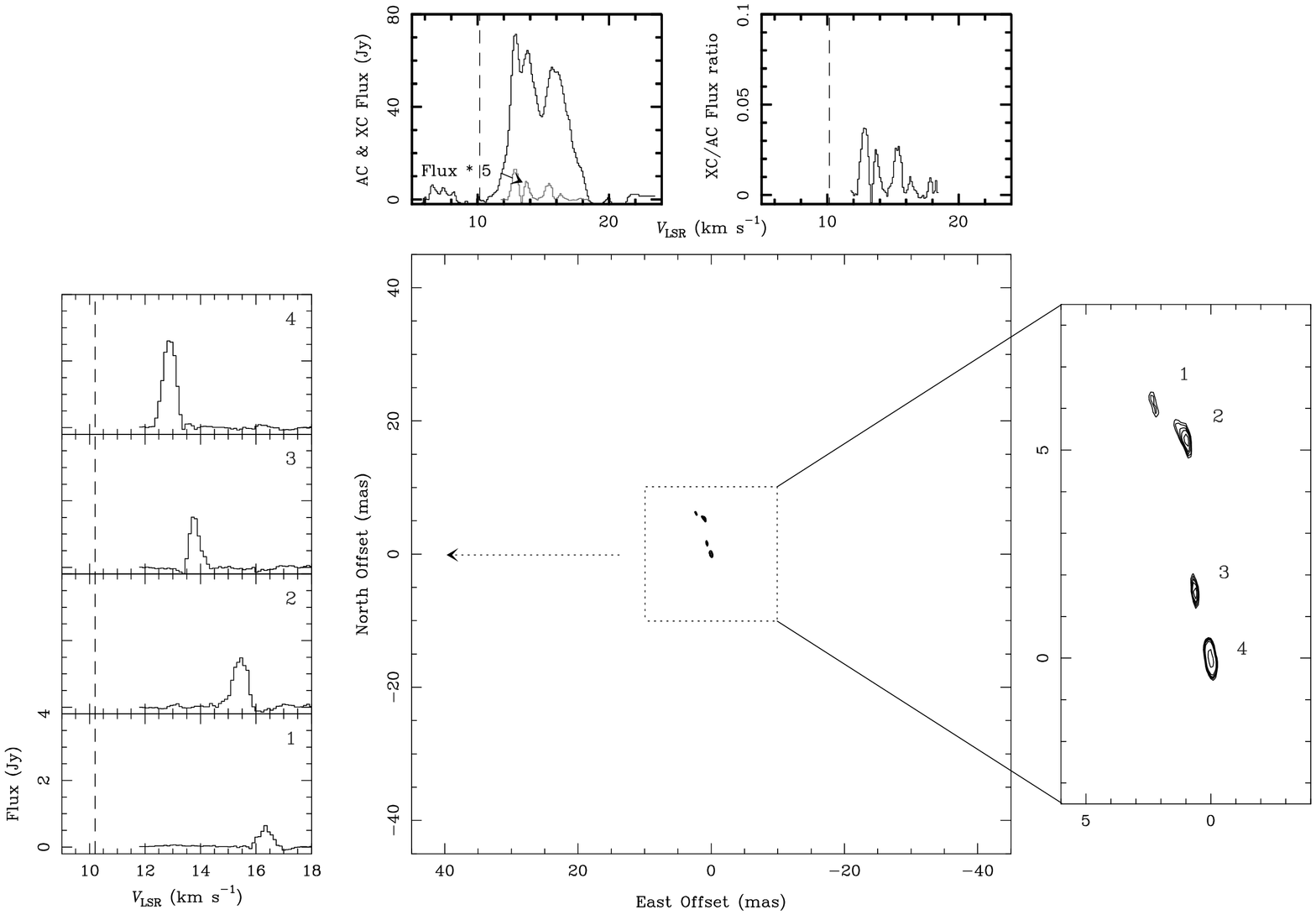}
      \caption{Same as Fig.\,\ref{IRCv1} for the $v$=2 $J$=2--1 SiO transition in $\chi$\,Cyg. 
The peak intensity of the map is 1.35 Jy beam$^{-1}$$\cdot$\,km\,s$^{-1}$ and the contour levels are 0.29,
 0.33, 0.37, 0.43, 0.52 and 1.05 Jy beam$^{-1}$$\cdot$\,km\,s$^{-1}$.}
         \label{chi3mmv2}
   \end{figure*}

\subsubsection{The $v$=2 $J$=2--1 transition}

 The first VLBA map of the $v$=2 $J$=2--1 $^{28}$SiO transition is presented in Fig.\,\ref{chi3mmv2}. 
The image is composed of four spots with velocities ranging from 12 to 17\,km\,s$^{-1}$ 
and flux intensities up to 2.5 Jy. For this maser, more than 95\% of the emission is lost in the 
interferometric map.

%
%

\section{Comparison of the different SiO transitions}

As noted previously, most of the properties of circumstellar SiO masers 
derived from VLBI studies have been obtained on \mbox{M-type} Mira variables
and, in particular, from the observations of the 7\,mm masers, mainly the $v$=1 $J$=1--0. 
However, one of the peculiarities of these stellar SiO masers is that they occur in a variety
of lines, with different frequencies and excitation energies. Therefore, the comparison
of the simultaneous observations harbors a crucial piece of information to understand the 
SiO maser phenomenon as well as to better constrain the theoretical models that 
try to explain the pumping of these masers.
To date, most of the VLBI multi-transitional studies of SiO masers have focused on the
comparison between $v$=1 and $v$=2 $J$=1--0 lines (at 7\,mm), while the comparison
of SiO masers at different frequency bands remains almost unexplored. 

In this section we compare the distribution of the observed $^{28}$SiO 
maser lines in IRC\,+10011 and $\chi$\,Cyg. 
We discuss the differences found between our observational results and what is predicted
 by the theoretical models, which are found to fail mainly in reproducing the differences 
observed between the 7 and 3\,mm masers. 
Finally, we explore the influence of the line overlaps between ro-vibrational transitions of
H$_2$O and $^{28}$SiO in the excitation, and localization in the envelope, of the SiO masers.

\subsection{The $v$=1 and $v$=2 $J$=1--0 masers}

\subsubsection{Previous results}
The first comparison of VLBI observations of the $v$=1 and $v$=2 $^{28}$SiO 
masers, using the KNIFE interferometer, was performed by Miyoshi et al. (\cite{miyoshi}), 
concluding that these two lines were spatially coincident. 
More accurate VLBA observations have demonstrated that there is a separation between the 
$v$=1 and $v$=2 ring structures of several milli-arcsec, with the $v$=2 maser line in most 
cases located inside the $v$=1 ring (Desmurs et al. \cite{desmurs1}, \cite{desmurs2}; see 
also Yi et al. \cite{yi}).
 Recently, Cotton et al. (\cite{cotton}) have furthermore demonstrated that this relative 
location of the two 7\,mm maser lines is independent of the optical phase of the star. 
We have found in our simultaneous observations a similar result for IRC\,+10011, but not for
 $\chi$\,Cyg as we will see in detail in the forthcoming subsections.

\begin{figure*}[!ht]
   \centering
 \includegraphics[width=0.97\textwidth]{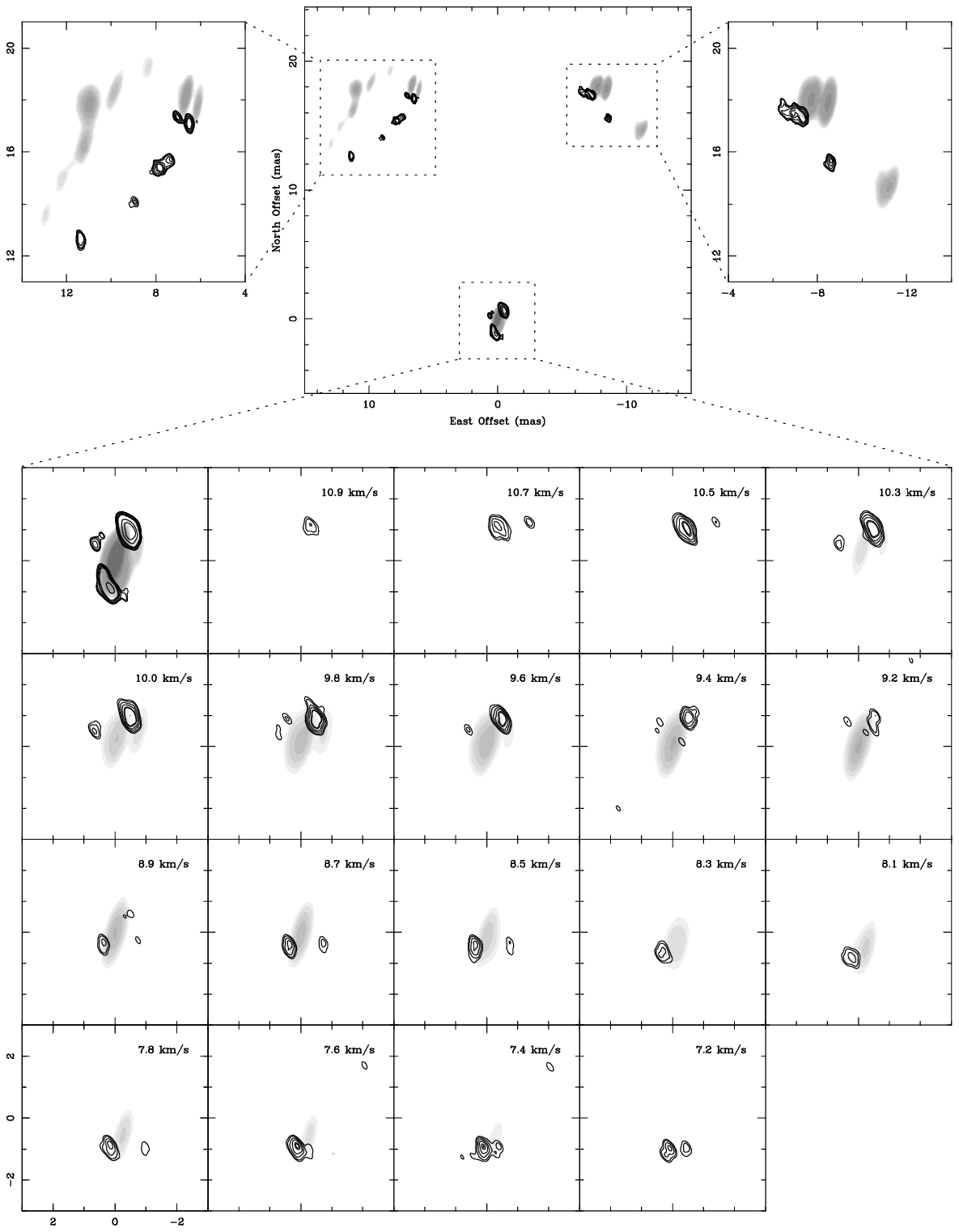}
      \caption{Comparison of the $v$=1 (greys) and $v$=2 (continuous contours) $J$=1--0 
transitions for IRC\,+10011. The zoom in plots are also displayed for the three masing 
regions. In the (0,0) region it has been represented the velocity averaged image (first panel) 
and the individual spectral emission channels covering the velocity range 10.9--7.2\,km\,s$^{-1}$. }
      \label{IRCcomp1}
   \end{figure*}

\subsubsection{IRC\,+10011}

The $v$=1 and $v$=2 $J$=1--0 SiO maps for IRC\,+10011 are shown in 
Figs.\,\ref{IRCv1} and \ref{IRCv2}. Clearly, the spatial distribution of the
two 7\,mm SiO lines is quite similar, in both cases the maser spots being grouped in 
three different regions. However, the emission of the $v$=2 line appears to be distributed 
in a region $\sim$\,10\% smaller than the $v$=1 one, \mbox{16--21 mas} vs. 19--24 mas, 
thus arising from a slightly inner region.

As we have mentioned in Section\,2, since the absolute position of the maps is lost in
the calibration process, their relative alignment is arbitrary. In this case, although 
assuming that the coincidence of the centroids of the two maps seems a reasonable choice, there
is an uncertainty in the centroid position of 2--3 mas. In Fig.\,\ref{IRCcomp1} we present
a possible alignment based on the fact that the spot 9 of the $v$=1 map shows
a velocity profile in between spots \mbox{9--11} of the $v$=2 line. If this is the right
alignment, one should conclude that there are no spatially coincident spots between
the two 7\,mm masers, in agreement with Desmurs et al. (\cite{desmurs2}). However, we 
must admit that other relative positionings are possible. Nevertheless, taking into account
that it is very unlikely to have spatially coincident spots with different velocity profiles, 
we can only match 2--3 spots at most (out of $\sim$~10 detected in both maps). Therefore, 
at least 70\% of the spots found in one transition do not appear in the other.

\subsubsection{$\chi$\,Cyg}

As noted in Section 3.2.2, only one feature was found in the $v$=2 
$J$=1--0 line, and so the comparison with the $v$=1 is not relevant for $\chi$\,Cyg,
since we can not make any reasonable assumption of the relative position between the
two maps. From the velocity profiles, the spot detected in the $v$=2 line could be
spatially coincident with spots either 2 or 4 in the $v$=1, or with none of them 
(see Figs.\,\ref{chiv1} and \ref{chiv2}).  
Nevertheless, since we have detected 6 spots in the $v$=1 $J$=1--0 maser, we should conclude 
that at least $\sim$\,80\% of the regions emitting in the $v$=1 have no counterpart in the 
$v$=2 transition. A similar conclusion can be reached just by comparing the total flux profiles.

\subsection{The $v$=1 and $v$=2 masers: $J$=1--0 vs. $J$=2--1}

\subsubsection{Previous results}

The only published comparison of quasi simultaneous VLBI observations of 7 and 
3\,mm SiO maser lines has been performed by Phillips et al. (\cite{phillips2}) in R Cas, which is an 
O-rich Mira-type variable. This source has a period of 430 days, and a mass loss rate 
of $\sim$\,1.1\,$10^{-6}$\,$M_{\sun}$\,yr$^{-1}$.
These authors observed the $v$=1 $J$=1--0 and $J$=2--1 masers of 
$^{28}$SiO, the former line being stronger and more complex at that epoch. More than 
ten $J$=1--0 spots delineate an incomplete circle. The $J$=2--1 line map consisted only 
of three spots, being two components possibly spatially coincident in both transitions.

\subsubsection{IRC\,+10011} 

\begin{figure*}[!ht]
   \centering
  \includegraphics[angle=-90, width=0.6\textwidth]{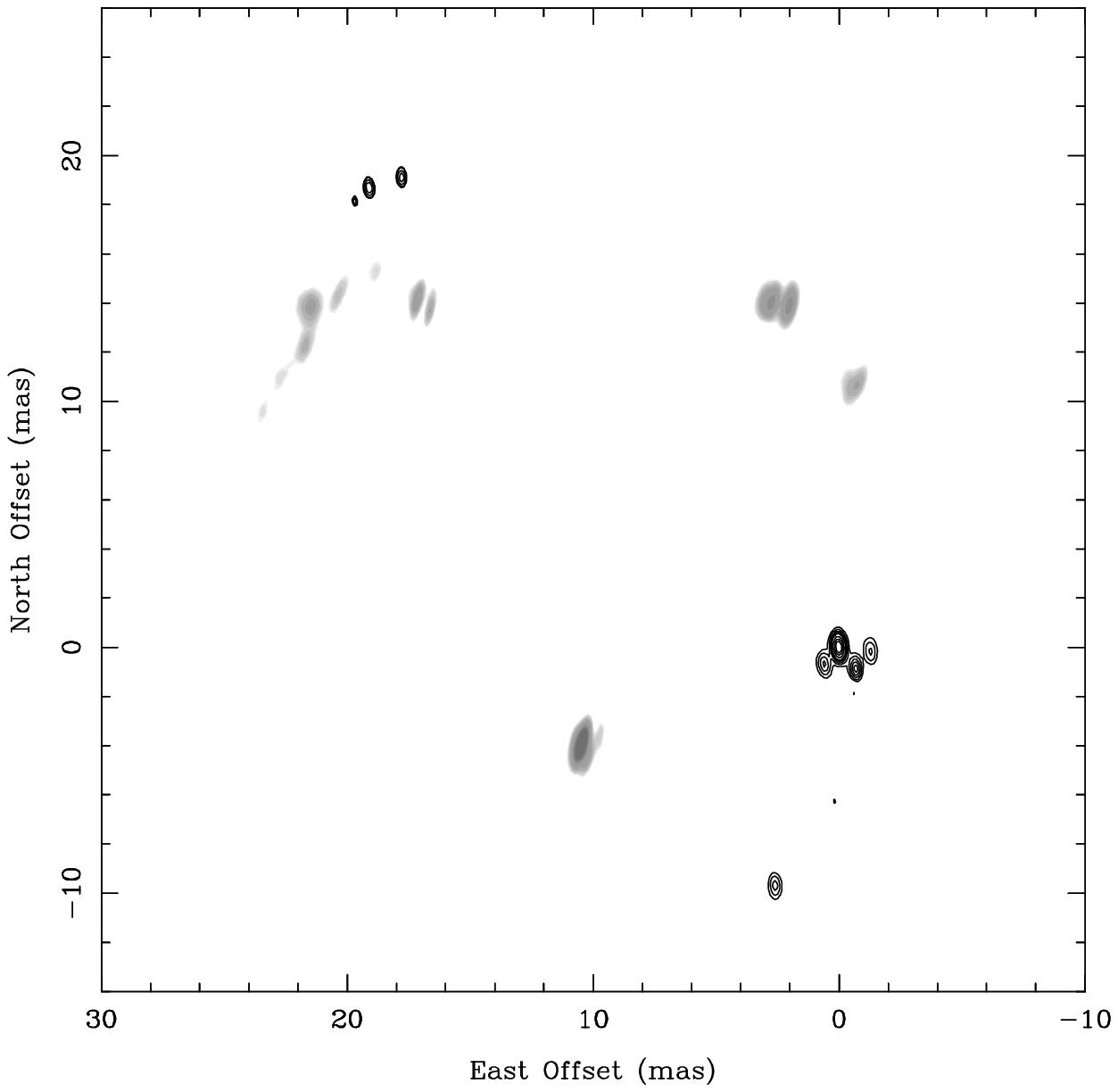}
      \caption{Comparison of the $v$=1 $J$=1--0 (greys) and $J$=2--1 (continuous 
contours) transitions for IRC\,+10011.}
      \label{IRC7y3}
   \end{figure*}

The maps obtained for the $v$=1 $J$=1--0 and $J$=2--1 in this source are displayed 
in Figs.\,\ref{IRCv1} and \ref{IRC3mm}. As it can be easily seen, the spatial 
distributions of both maser emissions are totally different. The separation between the two 
groups of spots in the 3\,mm line is much larger than the diameter of the maser emission ring 
in the $J$=1--0 line, 28--33 mas vs. 19--24 mas respectively. Assuming that the centroid 
of the emission is the same in both cases (see Fig.\,\ref{IRC7y3}), we can conclude that the 
$J$=2--1 originates from a shell located 50\% further away than the $J$=1--0, having no spots in 
common.
This is not surprising since the line profiles are not very similar (contrary 
to what happens between the two 7\,mm lines in this source). In fact, choosing a different 
alignment for the maps, it is impossible to obtain more than one coincident spot.

\subsubsection{$\chi$\,Cyg}

\begin{figure*}[!ht]
   \centering
 \includegraphics[angle=-90, width=0.6\textwidth]{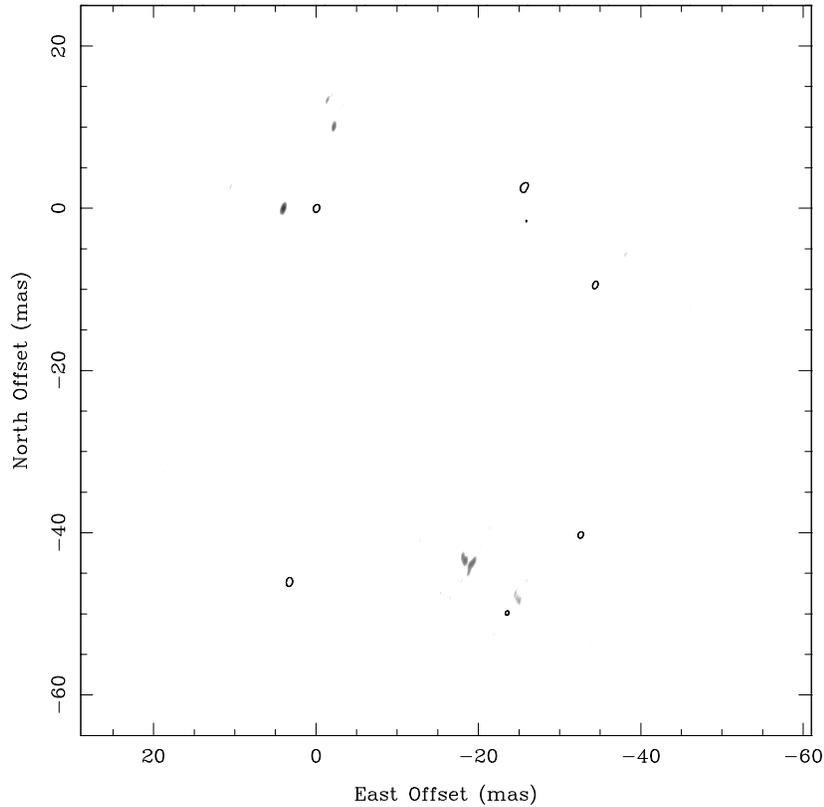}
      \caption{Comparison of the $v$=1 $J$=1--0 (greys) and $J$=2--1 (continuous 
contours) transitions for $\chi$\,Cyg.}
         \label{chicomp}
   \end{figure*}

When comparing the maps for the two $v$=1, $J$=1--0 and $J$=2--1, in $\chi$\,Cyg (see
Figs.\,\ref{chiv1} and \ref{chi3mm2}), we found a result somewhat different than that for
IRC\,+10011. Although the spots distribution is very different in the two transitions, 
the size of the emitting regions is comparable. This can be more easily appreciated 
in Fig.\,\ref{chicomp}, in which we have again assumed that the centroid of the two 
emissions is the same. 

The VLBA map of the $v$=2 $J$=1--0 maser is shown in Fig.\,\ref{chiv2}. We also present
the first VLBI map of the $v$=2 $J$=2--1 $^{28}$SiO maser in an AGB star in Fig.\,\ref{chi3mmv2}.
The $v$=2 $J$=2--1 maser emission is confined into an elongated region of 7\,mas,
displaying a clear velocity gradient of 3.5\,km\,s$^{-1}$ ($\sim$\,0.5\,km\,s$^{-1}$ per mas). 
These elongated structures are typical in other SiO maser sources (see the case of 
TX\,Cam, Diamond \& Kemball \cite{diamond2}). Due to this small spatial extent, it is very difficult 
to conclude a reasonable alignment for this map with respect to the $v$=1 line maps. 
This is also the case of the other $v$=2 transition, the $J$=1--0 line, which consists of a single spot.
Nevertheless, the comparison of the two emitting regions is not obvious since for the $J$=1--0 
there is only one maser component. However, its velocity distribution is quite similar to that 
of the spot number 4 of the $J$=2--1, and so both emissions could arise from the same region of 
the circumstellar envelope. 
However, where this region is located in the envelope is very difficult to say, since
the observed velocity range in the $v$=2 lines, $\sim$\,12--16\,km\,s$^{-1}$, is shared by several
spots of the $v$=1 masers.  

As we see, the available data to compare the 7 and 3\,mm masers are very scarce: this comparison has 
only been done in three sources, the two presented here, IRC\,+10011 and 
$\chi$\,Cyg, and R\,Cas by Phillips et al. (\cite{phillips2}). However, the results obtained can not be more 
disparate.
For the two ``normal'' stars, i.e. O-rich, we have one case, IRC\,+10011, in which the 
$J$=2--1 and $J$=1--0 $v$=1 masers arise from two totally different 
regions, with probably different physical conditions, whereas in R\,Cas Phillips et al. 
(\cite{phillips2}) claim that the masers originate from the same layer. For $\chi$\,Cyg, which on the 
contrary is an S-type star, the situation is more similar to R\,Cas. 
Nevertheless, we should note here that, as in the $v$=1 and $v$=2 masers, from 
the few existing observations, it seems that it is difficult to find maser spots emitting 
in both $v$=1 $J$=1--0 and $J$=2--1 lines.

\subsection{Comparison with theoretical models}

In an attempt to explain the existence and maintenance of SiO masers in the 
circumstellar envelopes of long period variable stars, several theoretical models have been proposed. 
Currently, two types of pumping mechanisms are considered, which are termed radiative 
(Kwan \& Scoville \cite{kwan}; Deguchi \& Iguchi \cite{deguchi}; Bujarrabal \& \mbox{Nguyen-Q-Rieu} 
\cite{bujarrabal1}; Bujarrabal \cite{bujarrabal4}, \cite{bujarrabal5}; Rausch et al. \cite{rausch})
 or collisional (Elitzur \cite{elitzur}; Lockett \& Elitzur \cite{lockett}; Doel et al. \cite{doel};
 Humphreys et al. \cite{humphreys1}, \cite{humphreys3}), depending on which is the 
primary mechanism responsible for this amplified emission. For radiative models, the energy of 
the maser pumping comes from the stellar radiation at 8\,$\mu$m, which corresponds 
to $\Delta v$=1 ro-vibrational SiO transitions. In collisional models, the kinetic energy of the gas 
pumps the masers by means of collisions with the H$_2$ molecules.

In all cases, the inversion of the levels in the maser transition is due to 
the same effect, the self-trapping of the photons in the $\Delta v$=1 ro-vibrational 
transitions when these lines are opaque (see Kwan \& Scoville \cite{kwan}). The main de-excitation path
 for molecules of SiO in the vibrational excited estates ($v$=1 and above) is the spontaneous 
decay via $\Delta v$=1 transitions ($v$=1--0, $v$=2--1, etc.). The corresponding de-excitation rate is 
proportional to 
$$ A_-/\beta_- + A_+ / \beta_+ $$
where $A_-$ and $A_+$ are the Einstein's A-coefficients for the spontaneous 
emission in the R- and P-branches, and $\beta_-$ and $\beta_+$ are the corresponding 
escape probabilities.
When the opacity is large, the escape probability in the envelope is given by 
the inverse of the opacity of the corresponding transition, which itself is proportional to 
the $J$ value of the upper state. In these conditions, the higher the $J$ 
value the more difficult for an SiO molecule to de-excite via $\Delta v$=1 
ro-vibrational transitions.
This results in a population inversion along the rotational ladder of the 
corresponding vibrationally excited state, and subsequently in a chain of maser transitions.

For this inversion mechanism to operate, the pumping rate must be independent of the $J$ 
quantum number. This is expected for collisional excitations, but also for radiative 
excitations if they are optically thin. 
The problem for the radiative models, of course, resides in producing at the same time $\Delta v$=1 
ro-vibrational transitions optically thin (excitation wise) and thick (de-excitation wise). 
This is solved assuming a large velocity gradient in the masing region or, more generally, a 
particular geometry in which the masers arise from a thin shell. Under these circumstances, the 
radial opacity in the $\Delta v$=1 lines can be optically thin, while the overall average along 
all directions is optically thick.
In spite of this difficulty, radiative models offer the advantage that the vibrational excitation
is much more probable than the corresponding collisional excitation. Moreover, it can 
naturally explain the thin rings of spots shown in VLBI observations, because the amplification
is tangential, and the tight correlation between the SiO maser intensity and the NIR stellar 
radiation (Bujarrabal et al. \cite{bujarrabal2}; Pardo et al. \cite{pardo}). On the other hand, 
collisional excitation operates under quite general conditions, provided that the temperatures 
exceed $\sim$\,1000\,K.

Because of the nature of the inversion mechanism, the physical conditions for 
the generation of SiO masers within the same vibrational state (i.e. $J$=1--0, $J$=2--1, etc.) are 
very similar regardless of the model. They occur when the corresponding $v \rightarrow 
v$--1 lines become optically thick, in a similar way for the different $J$ levels. On 
the contrary, for masers in different vibrationally excited estates (such as $v$=1 and 
$v$=2, etc.), the conditions must be naturally different, since it is difficult to have similar 
opacities for the $\Delta v$=1 ro-vibrational transitions of levels with very different 
excitation, simply because the population of the $v$=1 levels at $\sim$\,1780\,K will always
be higher than that of the $v$=2, for the expected temperature in the inner shells of the 
envelope (3400--1700 K). 
This is even more stringent for the radiative models, since to operate, the opacity of the ro-vibrational 
transitions must be $\sim$\,1, in order to be less than one in the radial direction but 
larger than one in average.

Because of these general results, simultaneous VLBI observations of several maser 
lines, both with similar and different excitation energies, in the same source offer the 
opportunity to better constrain the models, by comparing their predictions with the 
comparison of the maps obtained for the different lines. 


As described in previous sections, we have found a different result for the two sources studied.
In the case of $\chi$\,Cyg, the $J$=2--1 line is stronger than the $J$=1--0, and the emitting 
regions have comparable sizes. For the $v$=1 and $v$=2 $J$=1--0 lines, the $v$=1 maser is also 
stronger than the $v$=2. 
In IRC\,+10011, the differences found between the $v$=1 and $v$=2 $J$=1--0 lines are
smaller compared to those between the $v$=1 $J$=1--0 and $J$=2--1 transitions. In fact, the $J$=1--0 maser
emissions are quite similar, although not identical, with the $v$=1 being stronger and more extended. 
On the contrary, the $v$=1 transitions differ a lot, since we have obtained different angular sizes
 for both distributions, with the $J$=2--1 one being a $\sim$\,40\% larger. These results for 
IRC+\,10011 are incompatible with the theoretical predictions. 


 \begin{figure*}[!ht]
   \centering
 \includegraphics[width=0.9\textwidth]{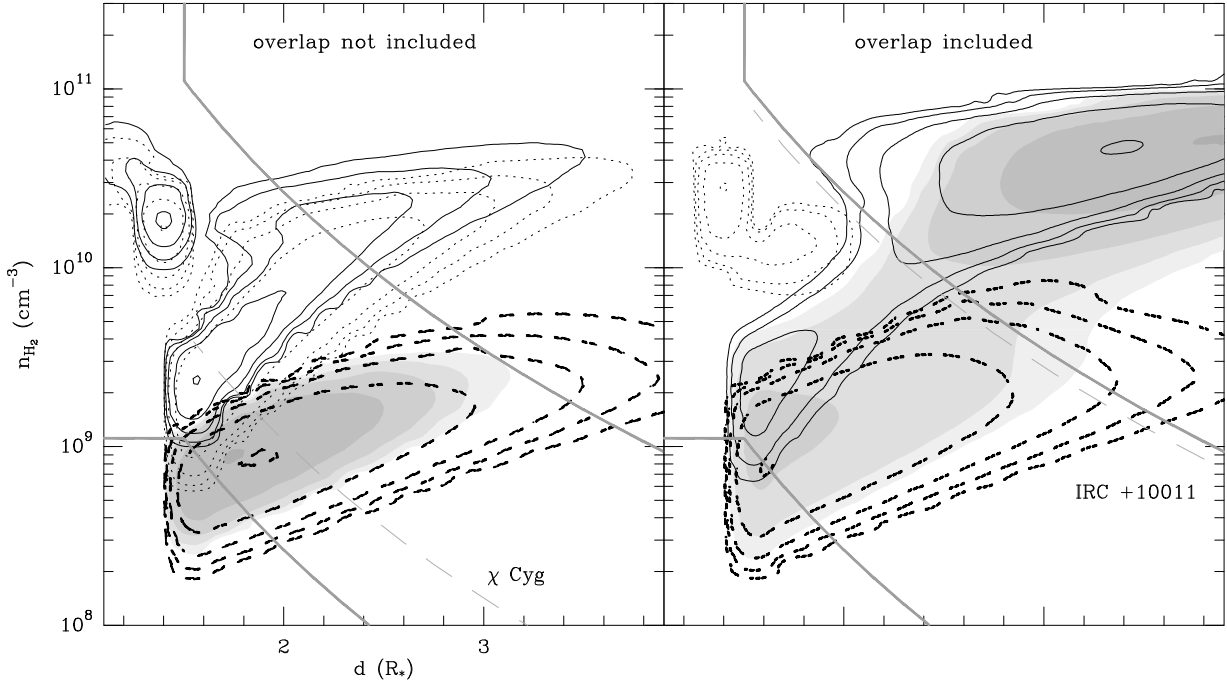}
      \caption{Model calculations of the excitation of the SiO molecule. Number of maser
photons as a function of the distance to the star and gas density, taking (right) and 
not taking into account (left) the effects of the line overlap between the ro-vibrational lines 
 $\nu_2$=0,12$_{7,5}$--$\nu_2$=1,11$_{6,6}$ of para-H$_2$O and $v$=1,$J$=0--$v$=2,$J$=1 
of $^{28}$SiO (greys: $v$=1 $J$=1--0, continuous contours: $v$=2 $J$=1--0, dashed contours:
 $v$=1 $J$=2--1, dotted contours: $v$=2 $J$=2--1). The thick grey lines mark the upper and 
lower limits of the mass loss rates in the sample of variable stars studied in BASCC, and 
the light grey dashed lines mark the mass loss rate of $\chi$\,Cyg and IRC\,+10011 respectively 
(see Table \ref{tab1}).} 
\label{modelox2}
\end{figure*}

\subsection{The effects of line overlaps}

One possible explanation for the different results found for IRC\,+10011
and $\chi$\,Cyg, when comparing the 7\,mm and 3\,mm SiO maser emissions 
(see Sect. 4.1 and 4.2), may reside in their different chemical 
composition, because of the effects of H$_2$O in the SiO excitation via the overlap 
of their ro-vibrational lines at 8\,$\mu$m.

It is well known that line overlap can strongly affect the pumping of maser
lines because their population inversion is very sensitive to small changes
in the population/de-population rates of the levels involved in the maser
transitions. In particular, for SiO masers, it was already proposed by Olofsson
et al. (\cite{olofsson}) that the overlap between the $v$=1,\,$J$=0\,--\,$v$=2,\,$J$=1 line of
$^{28}$SiO and the \mbox{$\nu_2$=0,\,12$_{7,5}$\,--\,$\nu_2$=1,\,11$_{6,6}$} line of
para-H$_2$O is responsible for the anomalous weakness of the
$v$=2 $J$=2--1 SiO maser in O-rich stars. These authors suggested this process in view
of the close frequencies of these two ro-vibrational transitions, less than 
1\,km\,s$^{-1}$, and of the relatively strong  $v$=2 $J$=2--1 lines
detected in S-type stars, for which lower H$_2$O abundances (compared to O-rich variables)
 are expected. This phenomenon was studied in more detail by Bujarrabal et al. 
(\cite{bujarrabal6}: BASCC hereafter), who found that introducing this line overlap in 
the model calculations, the observed properties of the $v$=2 $J$=2--1 maser of 
$^{28}$SiO in both O-rich and S-type stars could be explained.

The mechanism controlling the strength of the $v$=2 $J$=2--1 maser via line
overlap is very simple. Provided that the population of the $\nu_2$=2 states of
H$_2$O is dominated by collisions, the opacity and excitation of the 
$\nu_2$=0,\,12$_{7,5}$\,--\,$\nu_2$=1,\,11$_{6,6}$ of para-H$_2$O are larger
than those for the $v$=1,\,$J$=0\,--\,$v$=2,\,$J$=1 line of SiO. This results in an
enhanced net population of the $v$=2,\,$J$=1 SiO level, leading to an
 overthrow of the inverted population and the quenching off the 
$v$=2 $J$=2--1 maser (see BASCC for more details). For S-type stars, however, 
since they are characterized by a [O]/[C] ratio $\sim$\,1, 
the little Oxygen left in the gas phase after the formation of CO,
 should result in a lower H$_2$O abundance 
(no H$_2$O masers have been detected in S-type stars so far). One 
must conclude that the properties of SiO masers in S-type stars should be easier to 
reproduce by the standard models than those of O-rich stars, for which the effects 
of overlaps could be more important.

This also seems to be the case in our VLBI observations. For our M-type target,
IRC\,+10011, for which the effects of the line overlap can not be neglected (as
demonstrated by the non detection of the $v$=2 $J$=2--1 maser), the properties of
the maser spots distribution are in total disagreement with the predictions of standard models:
 similar properties for rotational transitions in the same vibrational state.
For $\chi$\,Cyg, in which the line overlap should not be as important as in M-type stars, 
(as demonstrated by the strength of its $v$=2 $J$=2--1 line), the properties of the maser 
spot distributions agree somewhat better with the expectations from the same standard models.

In order to verify the influence of this line overlap in the location of the
masers in the envelopes of \mbox{O-rich} sources, and to test whether this effect could
explain the results obtained for IRC\,+10011, we have performed new calculations using the
model developed by BASCC for SiO masers, which includes the
effects of the \mbox{$\nu_2$=0,\,12$_{7,5}$\,--\,$\nu_2$=1,\,11$_{6,6}$} line of para-H$_2$O.
We have studied the physical conditions under which the different SiO maser
lines arise, for a simplified envelope model, taking and not taking into
account the proposed line overlap. The characteristics of the numerical model,
as well as the physical parameters of the envelope are the same used in BASCC, 
except for the ratio between the ro-vibrational and pure-rotational collisional
coefficients of H$_2$O, which has been assumed to be 0.02 here (a value in the range studied 
by BASCC). The results of our calculations are summarized in Fig.\,\ref{modelox2}, 
where we plot the number of maser photons for the $v$=1 and 2, $J$=1--0 and \mbox{2--1} 
lines as a function of the distance to the star and gas density, introducing 
(right panel) and not introducing (left panel) the effects of the overlap.

As we can see, in the absence of overlap, the physical conditions required for
lines in the same vibrational state are very similar for the 7\,mm and 3\,mm masers
for both $v$=1 and $v$=2 transitions.
In addition, the $v$=2 lines require higher densities (and column densities)
than the $v$=1 masers, conditions that are very restrictive specially for envelopes
with low mass loss rates ($\sim$\,10$^{-7}$\,$M_{\sun}$\,yr$^{-1}$). 

When the overlap is included several changes occur:

\begin{description}
\item[a:] The $v$=2 $J$=2--1 maser is quenched, except for a very small
area in the space of physical conditions, in which the effect of the overlap is in fact
reversed because of the relatively low excitation temperature of the H$_2$O line (see
BASCC).
\item[b:] The $v$=1 $J$=2--1 line is barely affected by the overlap.
\item[c:] For conditions expected in the inner shells of the envelope,
especially at low mass loss rates, the $v$=2 \mbox{$J$=1--0} line requires lower 
densities (by a factor 2--3) to be masing.
\item[d:] The conditions for the $v$=1 $J$=1--0 to occur are now in between
those for the $v$=1 $J$=2--1 and $v$=2 \mbox{$J$=1--0}. This line is less coupled with the 
$v$=1 $J$=2--1 and becomes more similar to the $v$=2 $J$=1--0 line, as it is observed in IRC\,+10011.
This trend is more important for higher values of the ratio between the ro-vibrational
and pure-rotational collisional coefficients of H$_2$O: for values of this
parameter of 0.03, the $v$=1 and $v$=2 $J$=1--0 masing regimes become almost identical.

\end{description}

Although the modeling of the envelope in this preliminary study is very simple, the
results are nevertheless very encouraging, since they predict more or less what we have
observed in IRC\,+10011. However we believe that before investing more effort in this
theoretical modeling more observational data on the relative distribution of the SiO
maser lines, both in O-rich and S-type stars, are mandatory, as well as accurate 
estimations of the ro-vibrational collisional coefficients of H$_{2}$O with H$_{2}$.

\section{Conclusions}

We have performed high resolution observations of masers of the $^{28}$SiO 
molecule, in particular the rotational transitions $v$=1 $J$=1--0, $v$=1 $J$=2--1, $v$=2 
$J$=1--0 and $v$=2 $J$=2--1 in the regular variables IRC\,+10011 and $\chi$\,Cyg. The main 
goal of this work is to map each of these transitions so as to compare the 
spatial distribution of the different masers. Our results allow us to study in detail the 
differences between the observational results and the predictions of the theoretical 
pumping models, proving their validity in different types of sources.

We have presented in this paper the first VLBI maps for an S-type star, $\chi$\,Cyg: in particular, 
the $^{28}$SiO rotational lines $J$=1--0 and $J$=2--1 in the $v$=1 
and $v$=2 vibrational states. For the O-rich star, IRC\,+10011, we have mapped the 
$^{28}$SiO $v$=1 $J$=1--0 and $J$=2--1, and the $v$=2 $J$=1--0 maser emissions.


In IRC\,+10011, all the $^{28}$SiO maser transitions were detected and mapped except the $v$=2
$J$=2--1 line. When comparing the two 7\,mm maser emissions, we found that the $v$=2 is located 
in an inner region of the envelope, its angular extent being 10\% smaller than that of the $v$=2.
  On the contrary, for the $v$=1 $J$=1--0 and $J$=2--1 maser lines, the latter emission arises
in a layer $\sim$\,50\% farther away from the star.

For $\chi$\,Cyg, the $v$=1 $J$=1--0 and $J$=2--1 have comparable sizes, although their 
spatial distributions are different. On the other hand, the $v$=2 lines, $J$=1--0 and 
$J$=2--1, seem to originate in a similar region probably more compact and with more restrictive 
characteristics in the envelope.

Comparing our observational results with the different theoretical pumping mechanisms 
we obtain the following:

\begin{itemize}

\item[-] Standard models, collisional or radiative, have serious problems in explaining 
what is observed in IRC\,+10011: the $v$=1 and $v$=2 distributions being more similar than 
those of the $v$=1 $J$=1--0 and \mbox{$J$=2--1} transitions. This observational result 
contradicts what is predicted by all the theoretical studies, where the distributions of 
the $J$=1--0 and $J$=2--1 emissions appear to be similar. 
\item[-] However, in $\chi$\,Cyg, although the distributions of the $v$=1 $J$=1--0 and 
  $J$=2--1 transitions are different, as in IRC\,+10011, the sizes of the masing 
regions are comparable. For the $v$=2 $J$=1--0 and $J$=2--1 lines, it is difficult to establish 
the differences between the two emitting regions since, in the first transition only one feature 
was found.
\item[-] With respect to the $J$=2--1 and $J$=1--0 maser lines, a special effort must be made 
to study whether the observed relative spatial distribution of these emissions is similar 
to IRC\,+10011 or $\chi$\,Cyg. 
\item[-] The fact that almost none of the spots are coincident in all the transitions reveals 
that there may be additional maser competition effects that are not theoretically understood. 
We note that, since a considerable percentage of the emission is lost, the lack of coincidence 
could refer only to the brightest features.

\end{itemize}

To explain the discrepancies between theory and observations, we have proposed that the line overlap
between ro-vibrational H$_{2}$O and SiO lines is a process that strongly affects the conditions under 
which the masers occur. This effect is stronger in O-rich variables, such as IRC\,+10011, since they present
a higher abundance of H$_{2}$O in the envelope than S-type variables, such as $\chi$\,Cyg. The results are
 very promising: when the overlap is important, the $v$=1 and $v$=2 $J$=1--0 lines are strongly coupled
 and a significant difference appears between the $J$=1--0 and $J$=2--1 maser lines, as is observed 
in IRC\,+10011.   

We therefore conclude that a more extended study, from observational and theoretical points of view, 
will help to understand if the results for the SiO maser emissions obtained in IRC\,+10011 and $\chi$\,Cyg are 
produced systematically in other LPVs and to constrain both the pumping mechanisms and the importance of the
line overlaps.

\begin{acknowledgements}
 This work has been financially supported by the Spanish DGI (MCYT) under projects AYA2000-0927 and 
AYA2003-7584. We acknowledge with thanks the variable star observations from the AAVSO International Database 
contributed by observers worldwide and used in this research.
\end{acknowledgements}

\end{document}